\documentclass[aps,prb,twocolumn,groupedaddress,showpacs]{revtex4}

\usepackage{graphicx}

\begin{document}


\title{Checkerboard to Stripe Charge Ordering Transition in TbBaFe$_{2}$O$_{5}$}


\author{
D.K.~Pratt,$^{1,2}$
S.~Chang,$^{1,2}$
W.~Tain,$^{1,3}$
A.A.~Taskin,$^{4}$
Yoichi~Ando,$^{4}$
J.L.~Zarestky,$^{1,3}$
A.~Kreyssig,$^{1}$
A.I.~Goldman$^{1}$
and~R.J.~McQueeney$^{1}$
\\
\vskip 0.30cm
\centerline{}
\vskip 0.30cm
}

\affiliation{
\centerline{$^{1}$\footnotesize{Ames Laboratory, US DOE, Iowa State University, Ames, IA 50011, USA}}
\centerline{$^{2}$\footnotesize{NIST Center for Neutron Research, National Institute of Standards and Technology, Gaithersburg, Maryland 20899, USA}}
\centerline{$^{3}$\footnotesize{Oak Ridge National Laboratory, Oak Ridge, Tennessee 37831, USA}}
\centerline{$^{4}$\footnotesize{Institute of Scientific and Industrial Research, Osaka University, Ibaraki, Osaka 567-0047, Japan}}
}

\date{\today}

\begin{abstract}
A combined neutron and x-ray diffraction study of TbBaFe$_{2}$O$_{5}$ reveals a rare checkerboard to charge ordering transition.  TbBaFe$_{2}$O$_{5}$ is a mixed valent compound where Fe$^{2+}$/Fe$^{3+}$ ions are known to arrange into a stripe charge-ordered state below \textit{T}$_{\text{\textit{V}}}$  = 291 K, that consists of alternating Fe$^{2+}$/Fe$^{3+}$ stripes in the basal plane running along the \textbf{\emph{b}} direction.  Our measurements reveal that the stripe charge-ordering is preceded by a checkerboard charge-ordered phase between \textit{T}$_{\text{\textit{V}}}$ $<$ \textit{T} $<$ \textit{T}* = 308 K.  The checkerboard ordering is stabilized by inter-site coulomb interactions which give way to a stripe state stabilized by orbital ordering.
\end{abstract}


\maketitle
The charge-ordered (CO) state, where mobile electrons hopping between transition metal sites freeze into an ordered pattern, is a classic example of a localization-delocalization transition and has been under scrutiny for nearly a century.\cite{1}  Given its nature, the CO transition is typically a metal-insulator transition as well, and orbital selection of the ordered electrons controls the magnetic exchange pathways and, consequently, the magnetic properties of the material.  This phenomenon plays a critical role in the properties of many mixed valent oxides, such as cuprates, manganites, \cite{2} colbaltites, and ruthenates and consequently in the physics of colossal magneto-resistance, high temperature ferromagnetism, and unconventional superconductivity \cite{3,4}  In particular, an instability towards the formation of charge stripes of Cu$^{2+}$ ions in cuprate superconductors has been put forth as a possible driving force for unconventional superconductivity.\cite{5}

The CO phenomena can only occur in systems with a fractional formal valence.  In most cases, this fractional valence is introduced by chemical substitutions or the introduction of interstitial atoms.  However, this will also introduce quenched disorder as well as structural distortions due to steric effects that often complicate the physics. Studies of oxygen ion transport in pervoskite compounds, which explore ordering of oxygen vacancies, have revealed a rich variety of oxygen vacancy superstructures that can be realized and one of the rarer cases is the formation of ordered oxygen square bi-pyramids \cite{6} such as in \textit{R}Ba\textit{M}$_{2}$O$_{5}$ compounds (\textit{R} = rare earth, and \textit{M} = Mn, Fe, Co) \cite{7}. The formal valence of the metal ion is 2.5+, making them naturally mixed valent without the introduction of disorder, similar to the situation found in the classic charge-ordering system Fe$_{3}$O$_{4}$.\cite{1}  For \textit{M} = Fe, a charge-ordering (Verwey) transition occurs  (below \textit{T}$_{\text{\textit{V}}}$) where alternating stripes of Fe$^{2+}$ and Fe$^{3+}$ form and run along the crystallographic \textbf{\emph{b}}-direction (see Fig. 1).  In the CO state, the asymmetric orbital filling related to double occupancy of the \textit{d}$_{xz}$ orbital in the Fe$^{2+}$ valence state leads to a cooperative Jahn-Teller distortion and an orthorhombic structure.  Additionally, the half filling of the highest energy \textit{d}$_{z2}$ and \textit{d}$_{x2-y2}$ orbitals is consistent with the observed G-type antiferromagnetic (AFM) structure that persists above and below \textit{T}$_{\text{\textit{V}}}$ (Fig. 1).   Also, \textit{R}BaFe$_{2}$O$_{5}$ systems display a sensitivity of the direct magnetic exchange along the \textbf{\emph{c}}-axis which changes sign at the CO transition indicating an especially intimate connection between charge, orbital, and magnetic order and subsequent transport properties.

\begin{figure}[]
\includegraphics[width=1.0\columnwidth]{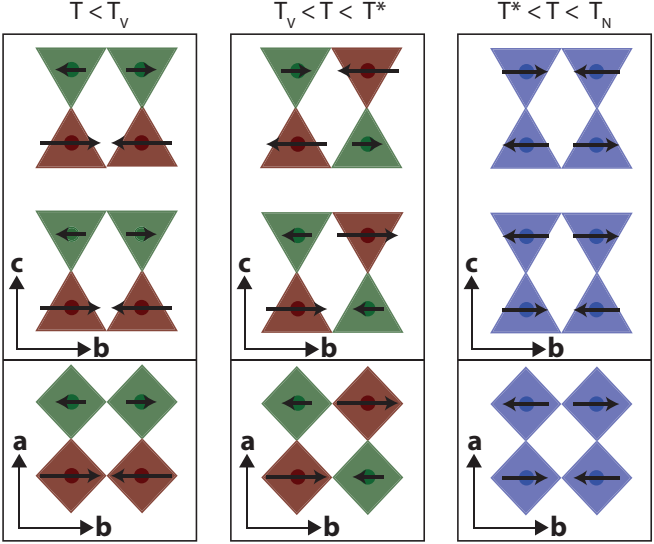}
\caption{Charge and spin ordering structures for the stripe CO phase for \textit{T} $<$ \textit{T}$_{\text{\textit{V}}}$ (left), checkerboard CO phase for  \textit{T}$_{\text{\textit{V}}}$ $<$ \textit{T} $<$ \textit{T}* (center), and mixed valent phase \textit{T}* $<$ \textit{T} $<$ \textit{T}$_{\text{\textit{N}}}$ (right) in both the \textbf{a}-\textbf{b} and \textbf{b}-\textbf{c} crystallographic planes. Triangles and squares correspond to different nominaly Fe ionic valences in the square pyramidal coordination; Fe$^{2+}$ (green), Fe$^{3+}$ (brown), and Fe$^{2.5+}$ (blue).  Arrows denote the magnetic moment of the respective ions.}
\label{mgraph2}
\end{figure}

	Woodward et al. explored this connection through a systematic study of \textit{R}BaFe$_{2}$O$_{5}$ (\textit{R} = Ho, Dy, Tb, Gd, Eu, Sm, and Nd) \cite{8,9,10,11,12}. For all compounds, a signature of another transition was observed above \textit{T}$_{\text{\textit{V}}}$ that is characterized by subtle anomalies in heat capacity, resistivity, magnetization, and changes in unit the unit cell volume.  This transition at \textit{T}* $>$ \textit{T}$_{\text{\textit{V}}}$, is described as a first attempt to stabilize an ordered arrangement of partially localized Fe$^{2.5+\epsilon}$ and Fe$^{2.5-\epsilon}$ ions (a first attempt to order charges as the system passes through the mixed valence (MV) states classified by Robin and Day \cite{13}).  However, diffraction studies on powder samples have not revealed evidence of any charge ordering associated with the so-called "premonitory" transition at \textit{T}*.

	In this paper, we report on a neutron and x-ray diffraction study of the Verwey transition and "premonitory" transition in a TbBaFe$_{2}$O$_{5}$ single crystal.  We show that, rather than being a premonitory transition, the phase existing between \textit{T}$_{\text{\textit{V}}}$ $<$ \textit{T} $<$ \textit{T}* is actually another CO phase where the iron valence orders in a checkerboard arrangement. Neutron scattering (NS) measurements of the magnetic structure confirm the expected modulation of the magnetic moment size as a consequence of the checkerboard CO state.  Thus, this compound presents a rare transition between two different charge ordered states that is likely driven by a competition between inter-site Coulomb interactions and orbital ordering.

High-quality TbBaFe$_{2}$O$_{5+x}$ single crystals have been grown at the Institute of Scientific and Industrial Research in Osaka Japan by the floating-zone technique using an infrared image furnace with two halogen lamps and double ellipsoidal mirrors.  A polycrystalline feed rod for the crystal growth was prepared by the solid-state reaction of Tb$_{4}$O$_{7}$, BaCO$_{3}$, and Fe dried powders. The crystal growth was performed at a constant rate of $0.5$ mm/h in flowing argon/hydrogen mixture. To obtain the stoichiometry of x=$0$, as-grown crystals were annealed in atmosphere of 93\% argon and 7\% hydrogen at $700$C for several days.

\begin{figure}[]
\includegraphics[width=1.0\columnwidth]{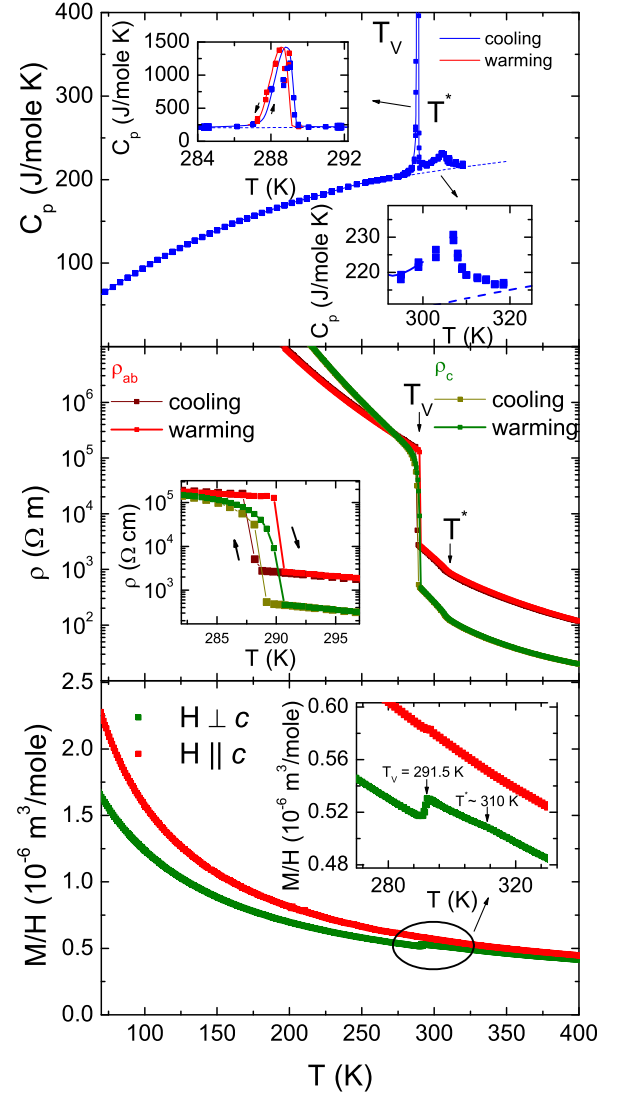}
\caption{Temperature dependence of heat capacity (a), resistivity (b), and magnetization (c) upon warming and cooling (Heat capacity measurements upon cooling were performed only around \textit{T}$_{\text{\textit{V}}}$.).  Resistivity measurements were taken along the in- and out-of-plane directions.  Magnetization measurements were taken with a 1 kG applied field both parallel and perpendicular to the \textbf{\emph{c}}-direction.}
\label{mgraph2}
\end{figure}

\vspace{-3em}
\begin{center}
\begin{table*}[]
\caption{Summary of superlattice propagation vectors} 
\hfill{}
\begin{tabular}{|c|c|c|c|} 
\hline
\hline 
             & \textit{T} $<$ \textit{T}$_{\text{\textit{V}}}$ & \textit{T}$_{\text{\textit{V}}}$ $<$ \textit{T} $<$ \textit{T}* & \textit{T}* $<$ \textit{T} $<$ \textit{T}$_{\text{\textit{N}}}$ \\ [0.5ex] 
\hline 
Magnetic & \mbox{\boldmath{$\tau$}}$_{\mbox{\footnotesize{G1}}}$ = (1/2, 1/2, 0) & \mbox{\boldmath{$\tau$}}$_{\mbox{\footnotesize{G2}}}$ = (1/2, 1/2, 1/2) & \mbox{\boldmath{$\tau$}}$_{\mbox{\footnotesize{G2}}}$ = (1/2, 1/2, 1/2)\\ 
         &                 & \mbox{\boldmath{$\tau$}}\textsuperscript{*} = (0, 0, 1/2)     &                  \\
\hline
Charge   & \textbf{q}$_{\text{stripe}}$ = (1/2, 0, 0)   & \textbf{q}$_{\text{checker}}$ = (1/2, 1/2, 0)   & None             \\ [1ex] 
\hline 
\end{tabular}
\hfill{}
\label{table:nonlin} 
\end{table*}
\end{center}

The proposed charge and magnetic structures of TbBaFe$_{2}$O$_{5}$ are shown in Fig. 1.  It was shown in powder diffraction studies \cite{9} that below the AFM transition at \textit{T}$_{\text{\textit{N}}}$ = $448$K the system exhibits an YBaCuFeO$_{5}$ type magnetic structure \cite{14} (Fig. 1 far right) and a valence of 2.5+ on the Fe sites.    Fig. 2 shows the results of heat capacity, resistivity, and magnetization measurements for temperatures in the vicinity of the Verwey transition at \textit{T}$_{\text{\textit{V}}}$ =  $291$K.  A first order transition occurs at \textit{T}$_{\text{\textit{V}}}$ as evident by the significant discontinuity in heat capacity [Fig. 2(a)], resistance along both the \textbf{\emph{a}}-\textbf{\emph{b}} plane as well as the \textbf{\emph{c}}-direction [Fig. 2(b)] and the less prominent anomalies in magnetization taken with a 1 kG (0.1 T) applied field parallel and perpendicular to the \textbf{\emph{c}}-direction [Fig. 2(c)].  All measurements shown in Fig. 2 were taken as the sample was both cooled and warmed through the transitions.  The insets in  Fig. 2(a), 2(b) and 2(c) show evidence of hysteresis associated with the Verwey transition.   The AFM structure is slightly modified below \textit{T}$_{\text{\textit{V}}}$ caused by a switch from AFM coupling between the double layers and the modulated magnetic moment magnitudes associated the Fe$^{2+}$ and Fe$^{3+}$ sites as is shown in Fig. 1 (far left).  Above \textit{T}$_{\text{\textit{V}}}$, the presence of a ferromagnetic double exchange interaction along \textbf{\emph{c}} proposed for GdBaFe$_{2}$O$_{5}$ \cite{15} is consistent with the dominant out of plane conductivity.  Finally, Figs. 2(a)-(b) clearly show evidence of the phase transition at \textit{T}* = $308$ K, whereas a feature in the magnetization data is much more subtle. These features at \textit{T}* do not show evidence of hysteresis and the transition appears to be second order in nature.

\begin{figure}[]
\includegraphics[width=1.0\columnwidth]{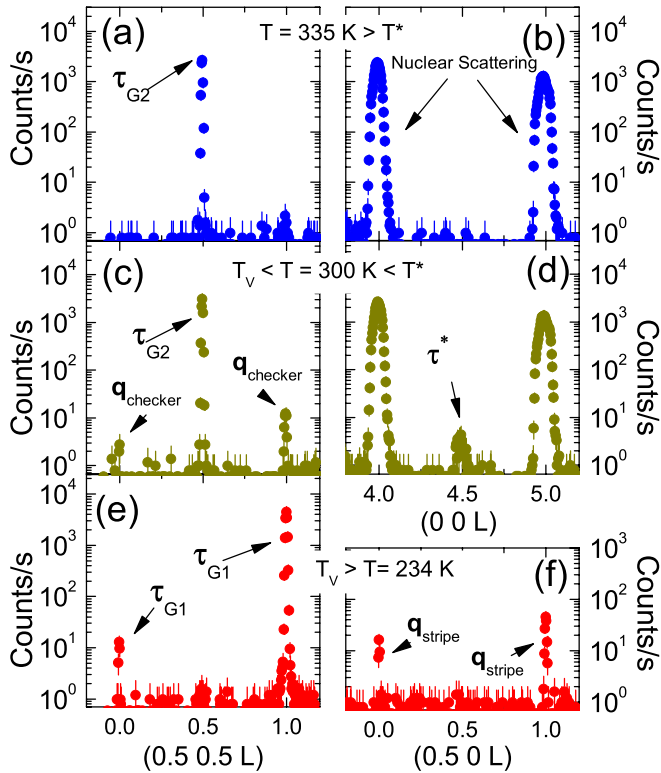}
\caption{Neutron scattering measurements taken (a)-(b) between \textit{T}$_{\text{\textit{N}}}$ and \textit{T}* showing the AFM ordering in the mixed valent phase at \mbox{\boldmath{$\tau$}}$_{\mbox{\footnotesize{G2}}}$, (c)-(d) between \textit{T}* and \textit{T}$_{\text{\textit{V}}}$ showing three characteristic propagation vectors in the {\textquotedblleft}premonitory{\textquotedblright} phase; \mbox{\boldmath{$\tau$}}$_{\mbox{\footnotesize{G2}}}$, \textbf{q}$_{\text{checker}}$, and \mbox{\boldmath{$\tau$}}\textsuperscript{*}, and (e)-(f) below \textit{T}$_{\text{\textit{V}}}$ showing \mbox{\boldmath{$\tau$}}$_{\mbox{\footnotesize{G1}}}$ and \textbf{q}$_{\text{stripe}}$ propagation vectors characterizing the stripe CO state.  The error bars indicate
a statistical error of one standard deviation.}
\label{mgraph2}
\end{figure}

	In order to explore the microscopic details of the charge and magnetic ordering in the {\textquotedblleft}premonitory{\textquotedblright} phase between \textit{T}$_{\text{\textit{V}}}$ and \textit{T}*, neutron diffraction measurements were performed at the High Flux Isotope Reactor (HFIR) at Oak Ridge National Laboratory.  The sample was placed in a cryo-furnace capable of reaching temperatures from 6 - 600 K and the collimator settings were 48$''$-40$''$-40$''$-68$''$.  Measurements were taken in the (\emph{HHL}) and (\emph{H}0\emph{L}) scattering planes in reciprocal lattice units associated with the tetragonal \textit{P4/mmm} crystallographic structure adopted by TbBaFe$_{2}$O$_{5}$ above \textit{T}$_{\text{\textit{N}}}$ = 448 K where a = 3.9 \AA and c = 7.6 \AA. \textbf{Q}-scans were performed along  (1/2,1/2,\textit{L}), (0,0,\textit{L}), and (1/2,0,\textit{L}) at \textit{T} = $265$, $295$, and $320$ K corresponding to temperatures in the stripe CO state (below \textit{T}$_{\text{\textit{V}}}$), in the premonitory state (\textit{T}$_{\text{\textit{V}}}$ $<$ T $<$ \textit{T}*), and in the mixed valent state (above \textit{T}*).  A summary of the results is shown in Fig. 3(a) - (f).  Fig. 3(a) shows a magnetic Bragg peak with propagation vector \mbox{\boldmath{$\tau$}}$_{\mbox{\footnotesize{G2}}}$ = (1/2,1/2,1/2) associated with the AFM structure in the mixed valent phase of TbBaFe$_{2}$O$_{5}$ above \textit{T}*.  Here the subscript $'$G$'$ refers to G-type magnetic order within the double layers while the $'$2$'$ indicates a doubling of the unit cell along \textbf{\emph{c}} due to ferromagnetic interactions between double layers\cite{9} (see Fig. 1).  Upon cooling below \textit{T}* into the premonitory phase, a new propagation vector \textbf{q}$_{\text{checker}}$ = (1/2,1/2,0) appears which, as we will show below, is the nuclear superlattice peak that characterizes the checkerboard CO state of the premonitory phase. While the system maintains the same magnetic propagation vector, \mbox{\boldmath{$\tau$}}$_{\mbox{\footnotesize{G2}}}$, Figs. 3(b) and (d) show that an additional magnetic modulation is observed at  \mbox{\boldmath{$\tau$}}\textsuperscript{*} = (0,0,1/2) below \textit{T}*.  Upon cooling into the stripe CO state below \textit{T}$_{\text{\textit{V}}}$, scans in Fig. 3(f) reveal superlattice peaks associated with stripe charge ordering at \textbf{q}$_{\text{stripe}}$ = (1/2,0,1). \cite{9}  We also find in Fig. 3(e) that the intensity at (1/2,1/2,1) increases by three orders-of-magnitude.  We now label  \mbox{\boldmath{$\tau$}}$_{\mbox{\footnotesize{G1}}}$=(1/2,1/2,1) as the magnetic propagation vector in the stripe CO state corresponding to G-type order within, and AFM coupling between, the double layers.  Table 1 includes a list of all CO and AFM propagation vectors.

NS measurements indicate long-range order in the premonitory phase with propagation vectors, (1/2,1/2,1/2), (1/2,1/2,0) and (0,0,1/2).  To verify whether these peaks are nuclear or magnetic, complimentary high-energy x-ray scattering measurements were performed at the Advanced Photon Source (APS) in the 6ID-D station.  Using $99.60$ keV and $129.6$ keV x-rays, diffraction patterns were recorded on a mar345 image plate detector system using a rocking technique described in ref. \cite{16}.  The sample was placed in a displex cryostat and inside a polyimide film  vacuum dome which had the capability of reaching $4$ - $800$ K. Figs. 4(a) and (b) show an image of the (\emph{H}0\emph{L}) plane below and above \textit{T}$_{\text{\textit{V}}}$, respectively.  As expected, superlattice peaks associated with the stripe charge ordering are observed below \textit{T}$_{\text{\textit{V}}}$ at \textbf{q}$_{\text{stripe}}$ = (\textit{n}/2,0,\textit{L}) with \textit{n} = odd disappear above \textit{T}$_{\text{\textit{V}}}$.  Figs 4(c)-(d) shows an image of the (\emph{HHL}) plane below and above \textit{T}$_{\text{\textit{V}}}$, respectively.  Fig. 4(d) shows that (1/2,1/2,\textit{L}) nuclear superlattice peaks appear between \textit{T}$_{\text{\textit{V}}}$ and \textit{T}* and can be associated with the CO charge vector, \textbf{q}$_{\text{checker}}$.  On the other hand, (1/2,1/2,\textit{L}) peaks do not appear in the x-ray patterns below \textit{T}$_{\text{\textit{V}}}$, indicating that they are purely magnetic reflections associated with AFM ordering propagation vector in the stripe phase (\mbox{\boldmath{$\tau$}}$_{\mbox{\footnotesize{G1}}}$).  Finally, we observe no evidence of (0,0,1/2) type reflections, confirming that \mbox{\boldmath{$\tau$}}\textsuperscript{*} is a purely magnetic modulation.  These combined x-ray and neutron results verify the existence of a transition from stripe to checkerboard CO in TbBaFe$_{2}$O$_{5}$.

The order parameters associated with various propagation vectors; \mbox{\boldmath{$\tau$}}$_{\mbox{\footnotesize{G2}}}$ = (1/2,1/2,1/2), \textbf{q}$_{\text{checker}}$ = (1/2,1/2,0),  \mbox{\boldmath{$\tau$}}\textsuperscript{*} = (0,0,1/2), \textbf{q}$_{\text{stripe}}$ = (1/2,0,1), and \mbox{\boldmath{$\tau$}}$_{\mbox{\footnotesize{G1}}}$ = (1/2,1/2,1) are constructed by fitting the peaks to a Gaussian line shape to obtain the integrated intensities, as shown in Fig. 5. Fig. 5(d) shows establishment of the stripe CO at \textbf{q}$_{\text{stripe}}$ = (1/2,0,1) with the corresponding switch of the magnetic propagation from \mbox{\boldmath{$\tau$}}$_{\mbox{\footnotesize{G2}}}$ = (1/2,1/2,1/2) [Fig. 5(a)] to \mbox{\boldmath{$\tau$}}$_{\mbox{\footnotesize{G1}}}$ = (1/2,1/2,1) [Fig. 5(c)].  The development of checkerboard CO with \textbf{q}$_{\text{checker}}$ = (1/2,1/2,0) is seen in both the x-ray and NS measurements in Fig. 5(c) between \textit{T}$_{\text{\textit{V}}}$ and \textit{T}*. The absence of x-ray intensity at (1/2,1/2, \textit{L} = 0 and 1) below \textit{T}$_{\text{\textit{V}}}$ [Fig. 5(c)] indicates that no crystallographic component exists at (1/2,1/2,\textit{L}) in the stripe ordered phase.  Lastly, Fig. 5(b) displays the additional magnetic modulation at \mbox{\boldmath{$\tau$}}\textsuperscript{*} = (0,0,9/2) characteristic of the checkerboard CO phase, whose origin is explained below.

\begin{figure}[]
\includegraphics[width=1.0\columnwidth]{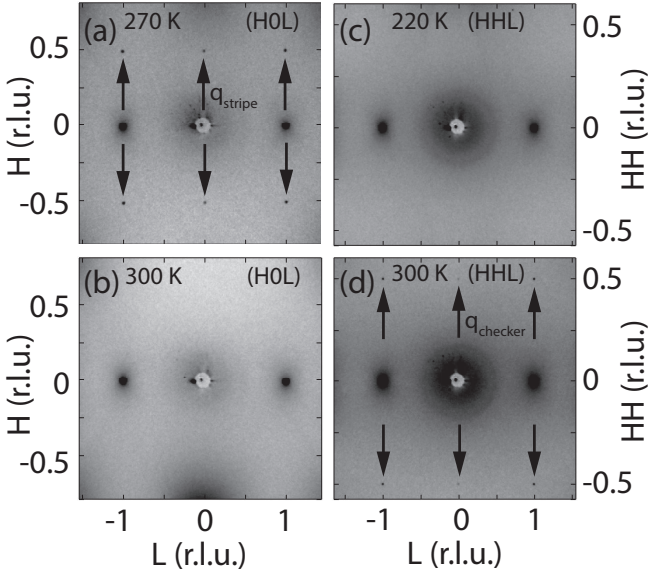}
\caption{X-ray diffraction data below \textit{T}$_{\text{\textit{V}}}$ in (a) the (\emph{H}0\emph{L}) plane showing \textbf{q}$_{\text{stripe}}$ and (c) the (\emph{HHL}) plane below \textit{T}$_{\text{\textit{V}}}$ showing no superlattice reflections.  Between \textit{T}* and \textit{T}$_{\text{\textit{V}}}$, panel (b) shows the disappearance of stripe ordering in the (\emph{H}0\emph{L}) plane, and panel (d) the appearance of checkerboard CO peaks at \textbf{q}$_{\text{checker}}$ in the (\emph{HHL}) plane.}
\label{mgraph2}
\end{figure}

\begin{figure}[]
\includegraphics[width=1.0\columnwidth]{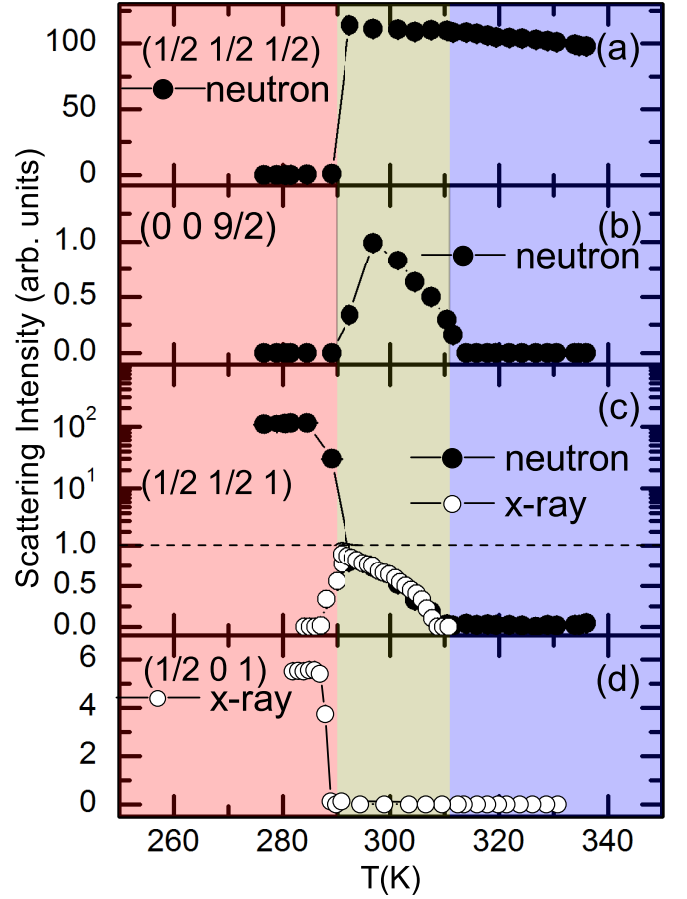}
\caption{Various charge and magnetic order parameters measured with neutron (filled circles) and x-ray scattering (open circles).  At \textit{T}$_{\text{\textit{V}}}$ = 291 K the appearance of stripe order associated with \textbf{q}$_{\text{stripe}}$ = (1/2,0,1) (d) results in switching of the mixed-valent magnetic structure from \mbox{\boldmath{$\tau$}}$_{\mbox{\footnotesize{G2}}}$ = (1/2,1/2,1/2) (a) to  \mbox{\boldmath{$\tau$}}$_{\mbox{\footnotesize{G1}}}$ = (1/2,1/2,1) (c).  The phase between \textit{T}$_{\text{\textit{V}}}$ $<$ \textit{T} $<$ \textit{T}* = 308 K, is characterized by checkerboard charge ordering at \textbf{q}$_{\text{checker}}$ = (1/2,1/2,1) in (c) and an associated moment modulation along the \textbf{\emph{c}}-direction in (b) with \mbox{\boldmath{$\tau$}}\textsuperscript{*} = (0,0,1/2).  Note the transition to a log intensity scale in the upper part of panel (c).  Scattering intensity error bars which represent statistical errors of one standard deviation are included but are smaller than symbols.}
\label{mgraph2}
\end{figure}

There are two possible ordered arrangements of iron valences which are consistent with \textbf{q}$_{\text{checker}}$=(1/2,1/2,0) as discussed in ref.\cite{9}.  The first scheme (model 1) is shown in the center of Fig. 1 corresponds to the checkerboard charge ordering observed also for YBaMn$_{2}$O$_{5}$ \cite{17}.   In model 1, the different sizes of Fe$^{2+}$ and Fe$^{3+}$ moments result in a net magnetization in each double layer (i.e. each double layer is ferrimagnetic).  This net magnetization switches direction in the next double layer, resulting in an additional magnetic modulation with a period of 2c corresponding to \mbox{\boldmath{$\tau$}}\textsuperscript{*}.  A second ordering scheme, model 2, (shown in Fig. 6 of ref. [9]) is composed of towers of nominally Fe$^{2+}$ and Fe$^{3+}$ ions running along the \textbf{\emph{c}}-direction, although no known material orders into this structure.  Nonetheless, a similar magnetic modulation also occurs in model 2 where the phase of the modulation is shifted along the \textbf{\emph{c}}-axis.  Similar magnetic modulations are absent in the mixed-valent phase because of the equivalent moments (due to the identical average valence) on each iron site, and absent from the stripe ordered state because the moments are compensated, as shown in Fig. 1.  Thus, the observation of \mbox{\boldmath{$\tau$}}\textsuperscript{*} strongly supports the checkerboard CO and serves as an effective order parameter as indicated in Fig. 4(b).

	Charge ordering driven entirely by a minimization of the inter-site Coulomb energy favors checkerboard CO according to model 1, where application of the Anderson condition requires that Fe$^{2+}$ ions are surrounded by Fe$^{3+}$ ions of opposite valence.  On the other hand, the extra electron on the Fe$^{2+}$ ion has a strong tendency to reside in the \textit{d}$_{xz}$ orbital as a consequence of the orthorhombic distortion found below \textit{T}$_{\text{\textit{N}}}$.\cite{7}  As localization of the extra electron into the \textit{d}$_{xz}$ orbital occurs below \textit{T}*, the system will eventually favor stripe CO as a way to minimize elastic strain.  In YBaMn$_{2}$O$_{5}$, charge localization into the \textit{d}$_{x2-y2}$ orbital of Mn$^{3+}$ is more symmetrical, and checkerboard CO is extremely stable.\cite{17}

	Transitions between two different CO states are extremely rare.  In La$_{0.5}$Sr$_{0.5}$NiO$_{4}$, a broad transition between incommensurate stripe order and checkerboard order is observed and discussed in terms of competing charge and spin interactions\cite{18}, while in Pr(Sr$_{x}$Ca$_{2-x}$)Mn$_{2}$O$_{7}$ ($0$ $<$ $x$ $<$ $0.45$), an incommensurate checkerboard to incommensurate stripe transition is observed and was connected to a competition between lattice distortion and charge/orbital stripe formation\cite{19}.  In TbBaFe$_{2}$O$_{5}$, \textit{T}* corresponds to a pure 2nd-order CO transition, whereas the stripe-to-checkerboard transition at \textit{T}$_{\text{\textit{V}}}$ is 1st-order and driven by orbital ordering.

This work was supported by the Division of Material Sciences and Engineering, Office of Basic Energy Sciences, U.S. Department of Energy (U.S. DOE). Ames Laboratory is operated for the U.S. DOE by Iowa State University under Contract No. DE-AC02-07CH11358. The work at the High Flux Isotope Reactor, Oak Ridge National Laboratory (ORNL), was sponsored by the Scientific User Facilities Division, Office of Basic Energy Sciences, U.S. DOE. ORNL is operated by UTBattelle, LLC for the U.S. DOE under Contract No. DEAC05-00OR22725.   Use of the Advanced Photon Source, an Office of Science User Facility operated for the US DOE Office of Science by Argonne National Laboratory, was supported by the US DOE under Contract No. DE-AC02-06CH11357.  We would also like to acknowledge the support from JSPS (NEXT program) and AFOSR (AOARD 124038).

\bibliography{TbBaFe2O5_IMP_Discovery}

\end{document}